\documentclass[twocolumn]{IEEEtran}

\usepackage{cite}
\usepackage{gensymb}
\usepackage{amsmath,amssymb,amsfonts}
\usepackage{bm}
\usepackage{amsthm}
\usepackage{textcomp}
\usepackage{graphicx}
\usepackage{algorithm}
\usepackage{algpseudocode}

\usepackage{multirow}
\usepackage{booktabs}
\usepackage{siunitx}
\usepackage{graphicx}
\usepackage{textcomp}
\usepackage{xcolor}
\usepackage{comment}
\usepackage{svg}
\usepackage{subcaption}

\usepackage[margin=16.5mm,top=19.5mm,bottom=26.6mm]{geometry}%margins;for printer deficiencies

\usepackage[normalem]{ulem} % option used to remove underline in journal name in references
\usepackage{cancel}
\usepackage{mathtools, cuted}

\usepackage{tikz}
\usetikzlibrary{shapes,arrows,positioning}
\usetikzlibrary{3d}

  {\proof}{\proofend}

\newcommand{\ME}{\mathbb{E}}

\newcommand{\vect}[1]{\boldsymbol{#1}}

\newcommand{\CC}{\mathbb{C}}
\newcommand{\RR}{\mathbb{R}}
\newcommand{\mc}[1]{\mathcal{#1}} % mathcal
\newcommand{\mb}[1]{\mathbf{#1}}
\newcommand{\herm}{^{\sf H}}

\newcommand{\jmathi}{\mathrm{j}}
\newcommand{\diff}{\mathrm{d}}

\newcommand{\qb}{{\bf b}}

\newcommand{\qe}{{\bf e}}

\newcommand{\qi}{{\bf i}}
\newcommand{\qj}{{\bf j}}
\newcommand{\qk}{{\bf k}}

\newcommand{\qr}{{\bf r}}
\newcommand{\qs}{{\bf s}}

\newcommand{\qy}{{\bf y}}

\newcommand{\qA}{{\bf A}}
\newcommand{\qB}{{\bf B}}

\newcommand{\qG}{{\bf G}}
\newcommand{\qH}{{\bf H}}
\newcommand{\qI}{{\bf I}}

\newcommand{\qQ}{{\bf Q}}

\DeclareMathOperator*{\argmin}{arg\,min}

\title{Distributed Continuous Aperture Arrays for Multiuser SWIPT}
\author{\IEEEauthorblockN{Muhammad Zeeshan Mumtaz, Mohammadali Mohammadi, Hien Quoc Ngo, and Michail Matthaiou}\\
\IEEEauthorblockA{{Centre for Wireless Innovation (CWI), Queen’s University Belfast, U.K.} \\
Email:\{mmumtaz01, m.mohammadi, hien.ngo, m.matthaiou\}@qub.ac.uk}
\thanks{

This work was supported by the UK Engineering and Physical Sciences Research Council (EPSRC) grant EP/X04047X/2 for TITAN Telecoms Hub. The work of  H. Q. Ngo
 was supported by the U.K. Research and Innovation Future
Leaders Fellowships under Grant MR/X010635/1, and a research grant from the Department for the Economy Northern Ireland under the US-Ireland R\&D Partnership Programme. This work was supported by the European
Research Council (ERC) under the European Union’s Horizon 2020 research
and innovation programme (grant agreement No. 101001331). }
\thanks{M. Z. Mumtaz is also with the College of Aeronautical Engineering, National University of Sciences \& Technology (NUST), Pakistan, (email: zmumtaz@cae.nust.edu.pk).}
}
\date{}

\begin{document}

\maketitle

\begin{abstract}
    This paper proposes a distributed continuous aperture array (D-CAPA) to support simultaneous wireless information and power transfer (SWIPT) to multiple information users (IUs) and energy users (EUs). Each metasurface supports continuous surface currents that radiate electromagnetic (EM) waves for information and energy transmission to the users. These waves propagate through continuous EM channels characterized by the dyadic Green’s function. We formulate a system power consumption (PC) minimization problem subject to spectral efficiency and energy harvesting quality‑of‑service (QoS) requirements, where the QoS requirements are derived under the equal power allocation (EPA) scheme. An efficient two-layer optimization algorithm is developed to solve this problem by optimizing the power allocation subject to the QoS violation penalties using augmented Lagrangian transformation. Our numerical results show that well‑optimized current distributions over each metasurface in the proposed D‑CAPA achieve up to $65\%$ and $61\%$ reductions in overall system PC compared to the EPA and co‑located CAPA (C‑CAPA) cases, while maintaining the same total aperture size and transmission power.
\end{abstract}
%\vspace{-0.5em}

\vspace{-1.5em}
\section{Introduction}
%------------------------------------

Continuous-aperture array (CAPA) transceivers synthesize continuous current distributions by replacing spatially discrete arrays (SPDAs) with a field-centric manifestation that emulates the classical Maxwell's EM propagation model \cite{Zhao:TWC2025,Yuanwei:WC:2025}. By manipulating these current distributions over the physical aperture, CAPA architectures unlock precise energy focusing and enhanced spatial degrees of freedom compared to the element-spaced arrays of comparable size  \cite{Wan:CLET:2023,Ouyang:TWC2025}. These unique functionalities translate into lower system PC as the transmitted energy is accurately concentrated on the network user locations with reduced spillover.

The EM field-based interpretation, that underpins the energy-focusing capability of CAPAs, enables us to pursue SWIPT in communication networks \cite{Qingxiao}. In particular, by effective utilization of these propagation characteristics, the current distributions over these metasurfaces can be exploited in a manner to achieve EM signal concentration for both EUs and IUs, while mitigating inter-user-interference (IUI) for IUs with aggressive sidelobe suppression \cite{Haiyang_Shlezinger}. 

Recently, CAPA technology has attracted significant research interest, including the hardware implementation of shared aperture metasurfaces \cite{Sazegar}. However, most of the related works have considered the case of a single CAPA surface for providing services in communication networks. In \cite{Zhaolin:TCOM_2025}, Wang \textit{et al.} conceived the optimal beamforming designs for a multiuser single-CAPA system with the aim of power minimization. The authors also demonstrated its superior performance against the SPDA variant. In \cite{Mengyu_Qian}, the authors availed of the calculus of variation (CoV) technique to handle the continuous nature of the energy distribution over a CAPA surface for a multigroup multicast network, while substantially improving the energy efficiency (EE). Similarly, metasurface-based holographic MIMO (H-MIMO) was considered in \cite{Qingxiao} for supporting integrated data and energy transfer by using finite-item Fourier series approximation for the continuous beamforming functions. However, these studies do not address the effects of distributed architectures on the PC reduction.

For the given overall aperture size and transmission power of the CAPA metasurfaces, we assess the performance gains of spatially D‑CAPA over its C‑CAPA counterpart. The main contributions of this paper are as follows:

\begin{itemize}
    \item We propose a D‑CAPA system that delivers SWIPT services using continuous transmission‑current signals over dyadic Green’s function‑based EM channels for both IUs and EUs. This architecture enables precise EM wave manipulation across spatially distributed metasurfaces, thereby enhancing both energy transfer efficiency and communication performance.
    \item A dual‑routine optimization framework is developed to address the PC minimization problem while ensuring compliance with the benchmark spectral‑efficiency (SE) and harvested‑energy (HE) QoS requirements. The proposed method employs the limited‑memory Broyden–Fletcher–Goldfarb–Shanno algorithm with bound constraints (L‑BFGS‑B) to optimize power allocation, while an augmented‑Lagrangian (AL) mechanism is used to manage QoS constraint violations.
    \item  Our numerical simulation results verify a substantial reduction in the collective PC of the considered optimized D-CAPA system  in comparison to the C-CAPA case. This performance enhancement is attributed to the precise beamforming capability achieved due to the focused energy zones over the continuous metasurfaces by the proposed optimization technique. 
\end{itemize}

% Main related references:
% \begin{itemize}
%     \item Main CoV based CAPA paper~\cite{Zhaolin:TCOM_2025,Mengyu_Qian}
%     \item Main CAPA based SWIPT paper ~\cite{Qingxiao}
%     \item Survey paper~\cite{Yuanwei:WC:2025}
%     \item Hardware CAPA implementation ~\cite{Sazegar}
%     \item CAP\cite{Jensen:AP:2008,Jeon:IT:2018,Migliore:JAP:2019,Wan:CLET:2023,Wan:TCOM:2023,Mikki:JAP:2023,Sanguinetti:TWC:2023,ouyang2024diversity}
%     \item CAP receiver~\cite{Ouyang:TWC:2025}
%     \item CAP and WPT~\cite{Haiyang_Shlezinger}
% \end{itemize}

\emph{Notations:} We use bold lower case and upper case letters to denote vectors and matrices, respectively. The superscripts $\!(\cdot)^{\rm{*}}\!$, $\!(\cdot)^{\rm{T}}\!$, and $\!(\cdot)^{-1}\!$  denote the conjugate, transpose, and inverse of a matrix, respectively; $\qI_N$ represents the $N\times N$ identity matrix; $\| \cdot \|$ returns the norm of a vector; $\ME\{\cdot\}$ represents the statistical expectation. Finally, $\nabla f(\cdot)$ performs the gradient operation on a function $f(\cdot)$.
% $\boldsymbol{1}_N$ denotes an all-one vector of size $N$;
% Finally, $\mathbb{E}\{\cdot\}$ denotes the statistical expectation.

%-----------------------------
% \vspace{-0.1em}
\section{System Model}
%--------------------------------
We consider a multi-metasurface system involving $S$ spatially distributed continuous aperture arrays which are indexed by $s\in\mathbb{S}=\{1,\ldots,S\}$, each having an effective aperture $\mc{S}_T^{(s)}$ with occupied area $A_T^{(s)}=\lvert{\mc{S}_T^{(s)}}\rvert= \bar{A}_T/S$, where $\bar{A}_T$ is the constant total aperture. It is further assumed that each metasurface is capable to synthesize any current distribution posed on its aperture. The downlink operation of these surfaces simultaneously serves $K = L + M$ users, represented by the union set $\mathcal{U} = \mathcal{D} \cup \mathcal{E}$. Here, $\mathcal{D} = \{1, \ldots, L\}$ denotes the IUs, while $\mathcal{E} = \{1, \ldots, M\}$ denotes the EUs. Now, the network topology is based on uniformly distributed CAPA surfaces with center locations $\mb{c}_s\in\RR^{3\times1}$. Moreover, the $K$ users are uniformly distributed in the three-dimensional (3-D) space surrounding the distributed CAPA surfaces, which are located on the plane $z = 0$.
%-----------------------------
% \vspace{-0.15em}
\subsection{CAPA Transmission Signal and Power Constraints}
%-----------------------------
Each CAPA surface operates with continuous current signals which is the superposition of current distributions intended for each user stream \cite{Qingxiao}. For the CAPA surface $s$, this phenomenon can be mathematically represented as
%-----------------------------
% \vspace{-0.1em}
\begin{equation}
\bm{J}_{s}(\bm{u}, \bm{\Omega}) = \sum\nolimits_{j=1}^{K} \Omega_{sj} \bm{\theta}_{sj}(\bm{r}_j,\bm{u})\,x_{sj}, 
\label{eq:tx-signal}
\end{equation}
%-----------------------------
where $\Omega_{sj}\in\bm{\Omega}$ is the wave amplitude which serves as the corresponding PA coefficient, while $\bm{\Omega}$ denotes the matrix of all the PA coefficients; $\mb{x}_{s}=[x_{s,1},\ldots,x_{s,L}, x_{s,L+1},\ldots,x_{s,L+M}]^{\rm T}\in\CC^{K}$ denotes the transmission symbol vector following a unit-variance Gaussian distribution, while satisfying the condition $\ME\{\mb{x}_{s}\mb{x}_{s}\herm\}=\mb{I}_{K}$; $\bm{\theta}_{sj}(\bm{r}_j,\bm{u}), \,\forall\, \bm{u} \in \mc{S}_T^{(s)}$, is the beamformer attributed to the current density component for the modulation of symbol $x_{sj}$ intended for user $j$ located at position $\bm{r}_j \in \mathbb{R}^{3 \times 1}$. Additionally, this composite transmitted signal is bounded by the available power budget at each CAPA surface as
%-----------------------------
% \vspace{-0.1em}
\begin{align}
     &\int_{\mc{S}_T^{(s)}} \lVert{\vect{J}_{s}(\vect{u}, \bm{\Omega})}\rVert^2\, \diff\vect{u} \leq P_{s}, \label{eq:perSurfacePower}
\end{align}
%-----------------------------
where $P_s=P_t/S$ is the maximum transmit power available at each surface out of the overall CAPA system power $P_t$.
%-----------------------------
\subsection{EM Channel Model}
\label{sec:channel}
%-----------------------------
The superimposed electric field $\vect{e}(\vect{r})$ at a position $\vect{r}=(r_x,r_y,r_z)\in\RR^{3\times 1}$ can be obtained as \cite{Wang:TWC:2025}
%-----------------------------
\begin{align}
    \vect{e}(\vect{r}) 
    = \sum\nolimits_{s=1}^S \int_{\mc{S}_T^{(s)}} \mb{G}_{s}(\vect{r},\vect{u})\, \vect{J}_{s}(\vect{u},\bm{\Omega})\, \diff\vect{u},
    \label{eq:eField}
\end{align}
%-----------------------------
where $\mb{G}_{s}(\vect{r},\vect{u})\in\CC^{3\times 3}$ is the dyadic free-space Green's function used to characterize the channel matrix from the surface aperture $\vect{u}\in\mc{S}_T^{(s)}$ to the user location $\vect{r}_k$, given by
%-----------------------------
\begin{align}
    \mb{G}_{s}(\vect{r},\vect{u}) 
    &= \frac{\jmathi\kappa Z_0}{4\pi} \frac{e^{\jmathi\kappa\lVert{\vect{p}}\rVert}}{\lVert{\vect{p}}\rVert}
    \Big[
    \big( \mb{I}_3 - \hat{\vect{p}}\hat{\vect{p}}\herm \big)
    + \frac{\jmathi}{\kappa\lVert{\vect{p}}\rVert}\big(\mb{I}_3-3\hat{\vect{p}}\hat{\vect{p}}\herm\big) \nonumber\\
    &\quad
    + \frac{1}{(\kappa\lVert{\vect{p}}\rVert)^2}\big(\mb{I}_3-3\hat{\vect{p}}\hat{\vect{p}}\herm\big)
    \Big],
    \label{eq:dyadic}
\end{align}
%-----------------------------
where $\vect{p}=\vect{r}-\vect{u}$, with the unit vector defined as $\hat{\vect{p}}=\vect{p}/\lVert{\vect{p}}\rVert$, $\kappa=2\pi/\lambda$ is the characteristic wave number with $\lambda$ being the carrier wavelength, while $Z_0$ represents the free-space impedance. The three difference terms in the definition of the Green’s function relate to the far-field, the middle-field and the near-field EM radiation regions. Consider this function, the continuous EM channel from the surface $s$ to a certain user $k$ can be defined by using the co-polar combining components of the tri-polarized CAPA framework as 
\begin{equation}
    h_{sk}(\vect{u}) \triangleq \hat{\qi}^{\rm T}_{k} \bm{G}(\bm{r}_k,\vect{u})\hat{\qi}_{s} +\hat{\qj}^{\rm T}_{k} \bm{G}(\bm{r}_k,\vect{u})\hat{\qj}_{s}+ \hat{\qk}^{\rm T}_{k} \bm{G}(\bm{r}_k,\vect{u})\hat{\qk}_{s}, 
\end{equation}
where $\langle\hat{\qi}_{s},\hat{\qj}_{s},\hat{\qk}_{s}\rangle$ and $\langle\hat{\qi}_{k},\hat{\qj}_{k},\hat{\qk}_{k}\rangle$ are unit vectors for the CAPA transmission surfaces and reception antennas, respectively. 
% (\textit{Extended the Mengyu Qian structure from uni-polarization to co-polar tri-polarization}).

% \section{Directions, Surface Equalization, and Benchmarks}
% \label{sec:directions}
% \hrule
\vspace{-1.2em}
\subsection{Beamforming Design}
The continuous-aperture formulation allows us to synthesize user-specific current distributions associated to the beamformer $\bm{\theta}_{sj}(\bm{r}_j,\bm{u})$ over the span of the continuous EM channels $h_{sk}(\bm{u})$. Mathematically, these beamformers are defined as $\bm{\theta}_{sj}(\bm{r}_j,\bm{u}) \triangleq {\theta}_{sj}(\bm{u}) \hat{\bm{p}} $ with normalized coefficients given 
%-----------------------------
\vspace{-0.2em}
\begin{align}\label{eq:beamformer_def}
\theta_{sj}(\bm{u})
&= \frac{\sum_{k=1}^K b_{jk}\,h_{sk}(\bm{u})}{\sqrt{\qb_j^{\!H}\qA^{(s)}\qb_j}},
\end{align}
%-----------------------------
where $\qA^{\!(s)}\!\!\triangleq\![\qA^{\!(s)}_{kk'}]_{k,k'} \!\in\!\CC^{K\!\times\! K}$ denotes the full-rank channel correlation matrix whose individual elements encode the correlation of  channel realizations induced on the CAPA surface, defined as \cite[Eq. (6)]{Ouyang2:TWC2025}  
%----------------------------- 
\begin{equation}
\qA^{(s)}_{kk'}\ \triangleq\ \int_{\mc{S}_T^{(s)}} h_{sk} (\bm{u})\,h_{sk'}^{*}(\bm{u})\,\diff\bm{u} .
\label{eq:chan_correlation}
\end{equation}
%----------------------------- 
Note that, $\qb_j$ in \eqref{eq:beamformer_def} represents the precoding vectors, which can be designed by utilizing linear techniques, i.e., regularized zero-forcing (RZF) and maximum ratio transmission (MRT) for the IUs and EUs, respectively. For RZF precoding, we extract the IU block from the composite channel correlation matrix as $\qA_{{L}} \subset \sum^{S}_{s=1}\qA^{(s)} \in \CC^{L \times L}$, which is used to construct the matrix $\qB_{L}=(\qA_{L}+\alpha_{\rm ZF} \qI_{L})^{-1}$, whose column elements are used as IU precoding coefficients $b_j$. The regularization factor $\alpha_{\rm ZF}>0$  trades between conventional ZF and MRT functionalities. On the other hand, the EUs employ the MRT technique derived from vector $\qb_{j}=\qe_{j}$, where $\qe_j$ is the canonical basis vector with $ e_{jm}=\delta_{jm}$. These MRT coefficients perfectly align the incident energy streams to the corresponding EM channels.
%-----------------------------
\vspace{-1em}
\subsection{Received Signal}
%-----------------------------
Each user can be assumed as a point in the 3-D radiation space of the CAPA transmission system, since the effective aperture of the receiver is negligible in comparison to the transmission framework, i.e., $A_R={\lambda^2}/{4\pi}\ll A^{(s)}_T$. The user $k$ located at the position $\qr_k$ receives the signal (from \eqref{eq:eField})
%-----------------------------
\begin{align}\label{eq:received_signal}
    \!y_k \!=\! \sum_{s=1}^S \!\Big(\Omega_{sk} x_{sk}\mathcal{J}_{sk}(\bm{G},\bm{\theta})\!+\! \sum_{j \neq k}^{K} \!\Omega_{sj} x_{sj}\mathcal{J}_{sj}(\bm{G},\bm{\theta})\!\Big)\!+\! n_k,
\end{align}
%----------------------------- 
 where the function $\mathcal{J}_{sk}(\bm{G},\bm{\theta})=\int_{\mc{S}_T^{(s)}} \mathcal{L}_{sk}(\bm{G},\bm{\theta})\, \diff\vect{u}$ is defined over the continuously differentiable function given as $\mathcal{L}_{sk}(\bm{G},\bm{\theta})=\hat{\qi}^{\rm T}_{k} \bm{G}(\bm{r}_k,\vect{u}){\theta}_{sk}(\bm{u})\hat{\qi}_{s} +\hat{\qj}^{\rm T}_{k} \bm{G}(\bm{r}_k,\vect{u}){\theta}_{sk}(\bm{u})\hat{\qj}_{s}+ \hat{\qk}^{\rm T}_{k} \bm{G}(\bm{r}_k,\vect{u}){\theta}_{sk}(\bm{u})\hat{\qk}_{s}$.
 Moreover, $n_k$ represents the EM noise with zero-mean Gaussian distributed entries $n_k\!\sim\!\mathcal{CN}(0,\sigma^2)$. 

%---------------------------------------
\subsection{Downlink SE and HE Analysis}
%-------------------------------------
For a given IU $l$, the corresponding signal-to-interference-plus-noise ratio (SINR) can be derived from \eqref{eq:received_signal} as
\begin{align}
    \Gamma_{l}(\qG,\bm{\theta},\bm{\Omega}) 
    = \frac{\sum_{s=1}^S \Omega^2_{sk}\lvert \mathcal{J}_{sk}(\bm{G},\bm{\theta})\rvert^2}{\sum_{s=1}^S  \sum_{j=1,j \neq k}^{K} \Omega^2_{sj} \lvert \mathcal{J}_{sj}(\bm{G},\bm{\theta})\rvert^2+\sigma^2_k}.
    \label{eq:SINR_NC}
\end{align}

Accordingly, the achievable SE (in bit/s/Hz) can be expressed as 
%-------
\begin{align}
    R_l=\log_2\!\left(1+\Gamma_{l}(\qG,\bm{\theta},\bm{\Omega})\right).
\end{align}
%---------------

% \subsection{Downlink EU RF Power}
% \label{sec:eu}
The power density incident on the antenna of a EU receiver $m$ can be evaluated as
%----------------------------- 
\begin{align}
    S(\qG,\bm{\theta},\bm{\Omega})=\dfrac{1}{2Z}\sum\nolimits_{s=1}^S \sum\nolimits_{j=1}^{K} \Omega^2_{sj} \lvert \mathcal{J}_{sj}(\bm{G},\bm{\theta})\rvert^2,
\end{align}
%----------------------------- 
where $Z$ is the wave impedance of the receiver. Using the Poynting's theorem, the RF power harvested at EU $m$ can be derived as
%----------------------------- 
\begin{align}
    \!\!Q_m (\qG,\bm{\theta},\bm{\Omega})\!&= \!\int_{\mc{S}_T^{(s)}}\!\! S(\qG,\bm{\theta},\bm{\Omega})  \,\hat{\bm{\eta}} \cdot\diff\vect{u}\nonumber\\
    &\hspace{-2em}=\! \frac{A_R\cos\varphi_m}{2Z} \sum\nolimits_{s=1}^S \sum\nolimits_{j=1}^{K}  \Omega^2_{sj} \lvert \mathcal{J}_{sj}(\bm{G},\bm{\theta})\rvert^2,
\end{align}
%----------------------------- 
where 
% $\mc{S}_R^{(m)}$ is the receiver aperture,
$\hat{\bm{\eta}}$ represents the normal vector to the receiver antenna, while $\varphi_m \in [0,\pi/2]$ denotes the including angle between $\hat{\bm{\eta}}$ and the Poynting vector at the $m$-th user location.

The practical implementation of EH corresponds to the non-linear harvesting energy (NL-HE) $Q^{\rm{NL}}_{m}$, which can be modeled using the following logistic function~\cite{Zeeshan:WCL:2025}
%----------------------------- 
\begin{align}
    &Q^{\rm{NL}}_{m}(\qG,\bm{\theta},\bm{\Omega}) = \frac{Q_{max}}{\varsigma\big(1+e^{-a(Q_m(\qG,\bm{\theta},\bm{\Omega})-b)}\big)} - \zeta,
    \label{eq:EH}
\end{align}
%-----------------------------
where $\varsigma={e^{ab}}/({1+e^{ab}})$
and $\zeta={Q_{max}}/{e^{ab}}$ with $a$ and $b$ are the EH rectifier parameters, while $Q_{max}$ denotes the saturation power of this rectifier circuit.
%-----------------------------
\section{Power Consumption Minimization Design}
%-----------------------------
\subsection{Optimization Problem Formulation}

The goal of the optimization problem is to minimize the overall system PC with respect to the beamformers, while ensuring that the SE and HE QoS requirements—derived from the EPA benchmark—are satisfied for all users associated with each CAPA surface. Mathematically, this optimization problem with continuous integral functions-based QoS constraints can be formulated as
%----------------------------- 
% \vspace{-0.1em}
\begin{subequations}
    \begin{align}
    \min_{\{\bm{\Omega}\}} 
    &\quad \sum\nolimits^S_{s=1}\sum\nolimits^{K}_{j=1} \Omega_{sj}^2 
    % \diff\bm{u} 
    \label{eq:optimization_prob}\\
    \text{s.t.}\quad 
    &\Gamma_l (\qG,\bm{\theta},\bm{\Omega})\ \geq\ \Gamma_l^{\rm EPA}, \ \forall \,l=1,2, \ldots, L, \label{eq:optimization_prob_C1}\\
    & Q_{m}(\qG,\bm{\theta},\bm{\Omega}) \geq Q_{m}^{\rm EPA}, \ \forall\, m =1,2, \ldots, M, \label{eq:optimization_prob_C2}\\
    &\sum\nolimits^{K}_{j=1} \Omega_{sj}^2 \leq P_{s}, \ \forall\, s=1,2, \ldots, S. \label{eq:optimization_prob_C3}
\end{align}
\end{subequations}
%----------------------------- 

For \eqref{eq:optimization_prob} and \eqref{eq:optimization_prob_C3}, it is assumed that the beamformers satisfy the condition 
$\int_{\mc{S}_T^{(s)}}\lVert\bm{\theta}_{sj}(\bm{r}_j,\bm{u})\rVert^2 \diff\bm{u}=1$. Moreover, the threshold SE ($\Gamma_l^{\rm EPA}$) and HE ($Q_{m}^{\rm EPA}$) in \eqref{eq:optimization_prob_C1} and \eqref{eq:optimization_prob_C2} have been computed for the  EPA case by using the PA coefficients $\Omega^{\rm EPA}_{sj}=P_t/(SK)$.  
%-------------------------
\vspace{-1em}
\subsection{Augmented-Lagrangian Optimization Framework}
%-------------------------
Considering the optimization problem in \eqref{eq:optimization_prob}, we leverage the AL assisted quasi-Newton method based L-BFGS-B approach \cite{Renbo:TSP2018,Jiaojiao:TSP2023}. This problem can be transformed into a finite sum minimization problem by employing the AL approach, which combines the continuous constraint functions in \eqref{eq:optimization_prob_C1} and \eqref{eq:optimization_prob_C2}. Mathematically, this AL transformation can be defined with respect to the PA coefficients choices ($\bm{\Omega}$) as
%-------------------------
\vspace{-0.2em}
\begin{align} \label{eq:Lagrangian}
\hspace{-0.7em}f(\bm{\Omega})\!
\triangleq& \!\sum^S_{s=1}\sum^{K}_{j=1}  \Omega_{s,j}^{2}
\!+
\underbrace{\bm{\lambda}_{\rm{SE}}^{\!\rm T}\bm{v}_{\rm{SE}}(\qG,\bm{\theta},\bm{\Omega})
\!+\!\frac{\rho}{2}\big\|\bm{v}_{\rm{SE}}(\qG,\bm{\theta},\bm{\Omega})\big\|^{2}}_{\text{IU SINR penalties}}\nonumber\\
&\hspace{-0em}\!+\!
\underbrace{\beta\!\left(\bm{\lambda}_{\mathrm{EH}}^{\!\rm T}\bm{v}_{\mathrm{HE}}(\qG,\bm{\theta},\bm{\Omega})
\!+\!\frac{\rho}{2}\big\|\bm{v}_{\mathrm{HE}}(\qG,\bm{\theta},\bm{\Omega})\big\|^{2}\right)}_{\text{EU HE penalties}},\!
\end{align}
%-------------------------
where, $\bm{v}_{\rm{SE}}(\qG,\bm{\theta},\bm{\Omega})=\bm{\Gamma}^{\rm EPA}-\bm{\Gamma}(\qG,\bm{\theta},\bm{\Omega})$ and $\bm{v}_{\rm{HE}}(\qG,\bm{\theta},\bm{\Omega})=\qQ^{\rm EPA}-\qQ(\qG,\bm{\theta},\bm{\Omega})$ represent the inequality residual vectors for IU SINR and EU HE \textbf{EPA} targets, which are arranged as $\bm{\Gamma}^{\rm EPA}$ and $\qQ^{\rm EPA}$, respectively; $\rho$ is the AL penalty; $\bm{\lambda}_{\rm SE}$ and $\bm{\lambda}_{\rm EH}$ are the Lagrange multipliers, and $\beta>0$ balances the IU and EU penalties considering the corresponding signal power differences. For the box constraints $0\le \Omega_{s,j}\le \sqrt{P_{s}}$, we use the L-BFGS-B technique to compute the projected quasi-Newton steps of the form
%-------------------------
\vspace{-0.2em}
\begin{equation} \label{eq:Newton_step}
    \bm{\Omega}_{t+1} = \Pi_{[0,\sqrt{P_{s}}]}\big(\bm{\Omega}_t + \alpha_t \bm{p}_t\big),
\end{equation}
%-------------------------
with search direction defined as $\bm{p}_t = -\qH_t \nabla f(\bm{\Omega}_t)$, while $\Pi_{[{0,\sqrt{P_s}]}}(\bm{\Phi}_t)= \argmin_{\bm{\Omega}\in \mathbb{R}^{+}} \{\lVert \bm{\Omega}_t-\bm{\Phi}_t \rVert\}$ projects the term $\bm{\Phi}_t=\bm{\Omega}_t + \alpha_t \bm{p}_t$ to the closest point within the feasible interval using Euclidean norm.

Here, $\nabla f(\bm{\Omega}_t) \in \mathbb{R}^{S \times K}$ represents the gradient of the Lagrangian function defined in \eqref{eq:Lagrangian} calculated for the step $t$. As both SE (${\Gamma}$) and HE ($Q$) are combinations of continuous integral functions with linear dependence on the PA coefficients, the Langrangian function in \eqref{eq:Lagrangian} can be differentiated using the chain rule. Each element of $\nabla f(\bm{\Omega})$ can be represented as partial derivative with respect to a particular PA variable $\Omega_{s,j}$ as
%-------------------------
\vspace{-0.25em}
\begin{align} \label{eq:grad_Lagrangian}
\dfrac{\partial f (\bm{\Omega})}{\partial \Omega_{s,j}} \!=& 2\,\Omega_{s,j}
\!-\!\sum\nolimits^L_{l=1} w_{l} \dfrac{\partial \Gamma_l(\qG, \bm{\theta}, \bm{\Omega})}{\partial \Omega_{s,j}}\!\nonumber\\
&-\!\sum\nolimits^M_{m=1} u_{m} \dfrac{\partial Q_m(\qG, \bm{\theta}, \bm{\Omega})}{\partial \Omega_{s,j}},
\end{align}
%-------------------------
where, $w_l\triangleq \lambda_{{\rm SE},l}+\rho\,v_{{\rm SE},l}$ and
$u_m\triangleq \lambda_{{\rm HE},m}+\rho\,v_{{\rm HE},m}$, while $v_{{\rm SE},l} \in \bm{v}_{{\rm SE}}(\qG, \bm{\theta}, \bm{\Omega})$ and $v_{{\rm HE},m} \in \bm{v}_{{\rm HE}}(\qG, \bm{\theta}, \bm{\Omega})$ are the inequality residuals for IU $l$ and EU $m$, respectively. The partial derivative of the SE component can be derived using the derivative quotient rule as

% Now, we evaluate the SE term in form of coherent and non-coherent cases: 

% For numerator: $\partial N = 2 \Omega_{s,l}\lvert \mathcal{J}_{sl}(\bm{G},\bm{\theta})\rvert^2$ for $j=l$.

% For denominator: $\partial D =2 \Omega_{s,j}\lvert \mathcal{J}_{sj}(\bm{G},\bm{\theta})\rvert^2$ for $j\neq l$

% By using quotient rule:

% \begin{align}
%     \dfrac{\partial \Gamma}{\partial \Omega_{s,j}}= \dfrac{D \partial N - N \partial D}{D^2}
% \end{align}
\vspace{-0.5em}
%-------------------------
\begin{align}\label{eq:partial_SE}
    \!\dfrac{\partial \Gamma}{\partial \Omega_{s,j}}\!\!=\!\!
    \begin{cases}
        &\!\!\!\!\!\!\dfrac{2 \,\Omega_{s,l} \lvert \mathcal{J}_{sl}(\bm{G},\bm{\theta})\rvert^2}{\sum_{s=1}^S  \sum_{j=1,j \neq l}^{K} \Omega^2_{sj} \lvert \mathcal{J}_{sj}(\bm{G},\bm{\theta})\rvert^2+\sigma^2_l},\quad \,\,\,\,\, j=l,\\
        &\!\!\!\!\!\! \dfrac{2\, \Omega_{s,j} \lvert \mathcal{J}_{sj}(\bm{G},\bm{\theta})\rvert^2 \sum_{s=1}^S \Omega^2_{sl}\lvert \mathcal{J}_{sl}(\bm{G},\bm{\theta})\rvert^2}{\big(\sum_{s=1}^S  \sum_{j=1,j \neq l}^{K} \Omega^2_{sj} \lvert \mathcal{J}_{sj}(\bm{G},\bm{\theta})\rvert^2+\sigma^2_l\big)^2}, \,\,\, j\neq l.
    \end{cases}
\end{align}
%-------------------------

Now, the partial derivative of the HE component can be given as
\begin{equation} \label{eq:partial_HE}
    \dfrac{\partial Q_m(\qG, \bm{\theta}, \bm{\Omega})}{\partial \Omega_{s,j}}= \frac{2 A_R\cos\varphi_m \Omega_{sj} \lvert \mathcal{J}_{sj}(\bm{G},\bm{\theta})\rvert^2}{2Z}. 
\end{equation}

Substituting the expressions in \eqref{eq:partial_SE} and \eqref{eq:partial_HE} in \eqref{eq:grad_Lagrangian}, we obtain the closed-form expression for the gradient of Lagrangian function in \eqref{eq:grad_Lagrangian_closed} at the top of the next page.

\begin{figure*}[ht]
    \begin{align} \label{eq:grad_Lagrangian_closed}
    \dfrac{\partial f(\bm{\Omega})}{\partial \Omega_{s,j}} \!=& 2\,\Omega_{s,j}
    \!- 2 \,w_{l} \Bigg(\dfrac{ \Omega_{s,l} \lvert \mathcal{J}_{sl}(\bm{G},\bm{\theta})\rvert^2}{\sum_{s=1}^S  \sum_{j=1,j \neq l}^{K} \Omega^2_{sj} \lvert \mathcal{J}_{sj}(\bm{G},\bm{\theta})\rvert^2+\sigma^2_l} +\!\sum^L_{l=1, j\neq l}  \dfrac{\Omega_{s,j} \lvert \mathcal{J}_{sj}(\bm{G},\bm{\theta})\rvert^2 \sum_{s=1}^S \Omega^2_{sl}\lvert \mathcal{J}_{sl}(\bm{G},\bm{\theta})\rvert^2}{\big(\sum_{s=1}^S  \sum\nolimits_{j=1,j \neq l}^{K} \Omega^2_{sj} \lvert \mathcal{J}_{sj}(\bm{G},\bm{\theta})\rvert^2+\sigma^2_l\big)^2}\Bigg)\!\nonumber\\
    &-\!\sum\nolimits^M_{m=1} \frac{2 u_{m}A_R\cos\varphi_m \Omega_{sj} \lvert \mathcal{J}_{sj}(\bm{G},\bm{\theta})\rvert^2}{2Z},
    \end{align}
    \hrule
    \vspace{-0.2em}
\end{figure*}

On the other hand, the matrix $\qH_t$ in \eqref{eq:Newton_step} represents a semi-positive definite limited-memory approximation for inverse-Hessian $\nabla^2 f(\bm{\Omega_t})^{-1}$ constructed from the last $q$ curvature pairs $\{(\qs_{i}, \qy_i)\}^{t-1}_{i=t-q}$ in order to avoid intense computational requirements for second order derivatives. This approximation evaluates the BDFS updates using the following relationship \cite[Eq. (6)]{Renbo:TSP2018}, 
%-------------------------
\begin{equation}
    \qH_{t+1}=
\bigg(\qI_q-\dfrac{\qs_t\qy_t^{\rm T}}{\qy_t^{\rm T} \qs_t}\bigg)\qH_t\bigg(\qI_q-\frac{\qy_t\qs_t^{\rm T}}{\qy_t^{\rm T} \qs_t} \bigg)
+\dfrac{\qs_t\qs_t^{\rm T}}{\qy_t^{\rm T} \qs_t},
\end{equation}
%-------------------------
where, $\qs_t \triangleq \bm{\Omega}_{t+1}-\bm{\Omega}_{t}$ and $\qy_t \triangleq \nabla f(\bm{\Omega}_{t+1})-\nabla f(\bm{\Omega}_{t})$ are the differentials of the successive PA updates and the gradients of their Lagrangian function, which is calculated by using the above closed-form expression in \eqref{eq:grad_Lagrangian_closed}.

\vspace{-0.4em}
%-------------------------
\subsection{Dual-Routine Optimization Algorithm}
%-------------------------
Now, we discuss the dual-routine optimization structure which computes the PA coefficient updates ($\bm{\Omega}$) using the L-BFGS-B method (inner iteration) and the Lagrange multiplier updates for AL-function (outer iteration), alternatively.  
%-------------------------
\subsubsection{L-BFGS-B Update Routine}
%-------------------------
For a particular inner iteration $t$, we generate the generalized Cauchy point ($\bm{\Omega^c_t}$) by following the steepest feasible direction $\nabla f(\bm{\Omega}_t)$ until the bounds $[0, \sqrt{P_s}]$ are reached. Then, we perform subspace minimization by applying L-BFGS-B two-loop recursion \cite[Algorithm 7.4]{nocedal2006numerical}, while using dual search directions ($\pm \bm{p}_t$) for the Newton step in \eqref{eq:Newton_step}. Additionally, the projected line-search function is defined as 
$\phi(\alpha_t)= f(\Pi_{[0,\sqrt{P_s}]}(\bm{\Omega}^c_t- \alpha_t \bm{p}_t)$, which is used to perform backtracking search to obtain line-search step size $\alpha_t>0$. Based on these updates, we evaluate the next PA estimate $\bm{\Omega}_{t+1}$ and utilize it to update the feasible curvature pair ($\qs_t, \qy_{t}$) and limited-memory Hessian $\qH_t$ based on the last $q$ curvature pairs stored temporarily. The initial choice of this Hessian matrix is set as $$\qH^{(0)}_t \triangleq \dfrac{\qs^{\rm T}_{t-1}\qy_{t-1}}{\lVert \qy_{t-1}\rVert^2}\qI_q.$$
It is important to note that the update step is required to satisfy the curvature condition $\qy_t^{\rm T} \qs_t>0$, which is regulated by AL penalty terms. The inner routine terminates as either the step size $\qs_t<\delta$ or gradient size $\qy_t<\delta$ with tolerance level $\delta$. 
%-------------------------
\subsubsection{AL Update Routine}
%-------------------------
During the iteration $u$, the outer routine updates the Lagrange multipliers $\bm{\lambda}_{\rm SE}$, $\bm{\lambda}_{\rm HE}$, subject to the QoS constraint violations. Mathematically, this can be represented as:
%-------------------------
\begin{subequations}
    \begin{align}
        \bm{\lambda}^{u+1}_{\rm SE}= \bm{\lambda}^{u}_{\rm SE}+\rho \bm{v}^{u}_{\rm{SE}},(\qG,\bm{\theta},\bm{\Omega}),\\
        \bm{\lambda}^{u+1}_{\rm SE}= \bm{\lambda}^{u}_{\rm HE}+\rho \bm{v}^{u}_{\rm{HE}}(\qG,\bm{\theta},\bm{\Omega}),
    \end{align}
\end{subequations}
%-------------------------
where the QoS violations for a particular iteration $u$ are calculated as $ \bm{v}^{u}_{\rm{SE}}(\qG,\bm{\theta},\bm{\Omega})=\max(\bm{\Gamma}^{\rm{EPA}}(\qG,\bm{\theta},\bm{\Omega})-\bm{\Gamma}(\qG,\bm{\theta},\bm{\Omega}),0)$ and $ \bm{v}^{u}_{\rm{HE}}(\qG,\bm{\theta},\bm{\Omega})=\max(\qQ^{\rm{EPA}}(\qG,\bm{\theta},\bm{\Omega})-\qQ(\qG,\bm{\theta},\bm{\Omega}),0)$. If the maximum of these violation criteria $v_{\max}=\max\{\|\bm{v}_{\rm SE}\|,\|\bm{v}_{\rm EH}\|\}$ reduces below the feasibility tolerance $\epsilon \geq 0$, the algorithm convergence is achieved. In the event of non-convergence, the AL penalty weight $\rho$ addresses the constraint violations adaptively over the successive AL-iterations. By adopting an aggressive policy, we double this parameter for the next AL-iteration $u+1$ as $\rho^{u+1}=2 \rho^{u}$. The steps of this process are presented in \textbf{Algorithm \ref{alg:PA_algorithm}}. 
%-------------------------

\begin{algorithm}[t]
\caption{AL + L-BFGS-B for PA optimization}
\begin{algorithmic}[1]
\State \textbf{Initialization:} Set iteration indices $t=0$, $u=0$; initial PA estimates $\bm{\Omega}_0=P_t/(SK)$; Lagrange multipliers $\bm{\lambda}_{\rm SE},\,\bm{\lambda}_{\rm HE}$; AL penalty $\rho>0$; IU/EU balancing factor $\beta>0$; feasibility thresholds $\delta=0$, $\epsilon>0$.
\State \textbf{AL Routine:}
\State \textbf{repeat}
\State \quad Define $f(\bm{\Omega})$ as in \eqref{eq:Lagrangian} with current $\bm{\lambda}_{\rm SE},\bm{\lambda}_{\rm HE},\rho$.
\State \quad Set inner iterate $\bm{\Omega}\gets\bm{\Omega}_u$. Initialize limited-memory pairs $\{(\qs_i,\qy_i)\}$ for L-BFGS-B approach.
\vspace{0.15em}
\State \quad \textbf{L-BFGS-B Routine:}
\State \quad \textbf{repeat}
\State \quad \quad Compute gradient $\nabla f(\bm{\Omega_t})$ using \eqref{eq:grad_Lagrangian_closed}.
\State \quad \quad Follow the steepest direction $\nabla f(\bm{\Omega}_t)$ to the bounds $[0,\sqrt{P_s}]$, obtaining generalized Cauchy point $\bm{\Omega}^{c}_t=\Pi_{[0,\sqrt{P_s}]}(\bm{\Omega}_t-\alpha\,\nabla f(\bm{\Omega}_t))$.
\State \quad \quad Apply the two–loop L-BFGS recursion to get the quasi-Newton direction $
\bm{p}_t = -\,\qH_t\,\big(\nabla f(\bm{\Omega}^{c}_t)\big).$
\State \quad \quad Define the projected line-search  function $\phi(\alpha_t)=f\!\left(\Pi_{[0,\sqrt{P_s}]}\!\big(\bm{\Omega}^{c}_t+\alpha_t\,\bm{p}_t\big)\right)$ and perform a backtracking search to get $\alpha_t>0$.
\State \quad \quad Update PA estimate: $\bm{\Omega}_{t+1}=\Pi_{[0,\sqrt{P_s}]}\!\big(\bm{\Omega}^{c}_t+\alpha_t \bm{p}_t\big)$.
\State \quad \quad Update curvature pair: \hfill 

$\qs_t=\bm{\Omega}_{t+1}-\bm{\Omega}_{t}$,
\quad $\qy_t=\nabla f(\bm{\Omega}_{t+1})-\nabla f(\bm{\Omega}_{t})$. 
\State \quad \quad \textbf{if}
$\qy_t^{\rm T}\qs_t>0$:
\State \quad \quad \quad Store the feasible curvature pair $(\qs_t,\qy_t)$
\State \quad \quad \quad Update the limited-memory inverse-Hessian $\qH_t$
\State \quad \quad \textbf{else} \textbf{continue}
\State \quad \quad Increment inner iteration index $t=t+1$.
\State \quad \quad \textbf{until} $\|\qs_t\|\leq \delta$ or $\|\qy_t\|\leq \delta$.
% \vspace{0.15em}
\State \quad Evaluate violations:\hfill

\!\!$\bm{v}_{\rm SE}(\qG,\bm{\theta},\bm{\Omega}_{u+1})=\max\!\big(\bm{\Gamma}^{\rm EPA}-\bm{\Gamma}(\qG,\bm{\theta},\bm{\Omega}_{u+1}),\,0\big)$,
% \State \quad \phantom{Evaluate violations}\hspace{3.7em}

\!\!$\bm{v}_{\rm HE}(\qG,\bm{\theta},\bm{\Omega}_{u+1})\!=\!\max\!\big(\qQ^{\rm EPA}-\qQ(\qG,\bm{\theta},\bm{\Omega}_{u+1}),\,0\big)$.
\State \quad Lagrange multiplier update:\hfill

$\bm{\lambda}_{\rm SE}\gets \bm{\lambda}_{\rm SE}+\rho\,\bm{v}_{\rm SE}$,\quad $\bm{\lambda}_{\rm HE}\gets \bm{\lambda}_{\rm HE}+\rho\,\bm{v}_{\rm HE}$.
\State \quad Increase penalty: $\rho\gets 2\rho$.
\State \quad Increment outer iteration index $u=u+1$.
\State \quad \textbf{until} $\max\{\|\bm{v}_{\rm SE}\|,\|\bm{v}_{\rm HE}\|\}\leq \epsilon$.
\vspace{0.15em}
\State \textbf{Return:} $\bm{\Omega}^{\star}=\bm{\Omega}_{u+1}$.
\end{algorithmic}\label{alg:PA_algorithm}
\end{algorithm}
%-------------------------
% \subsection{Computational Complexity}
%-------------------------

% Now, we evaluate the computational complexity of our proposed dual-routine optimization algorithm. 

\textbf{Complexity analysis:} The execution of the L-BFGS-B routine involves the calculation of finite difference gradient $\nabla f(\bm{\Omega}_t)$ and  limited memory Hessian matrix $\qH_t$, incurring per-iteration complexity of $\mathcal{O}(S K^2)$ and $\mathcal{O}(qSK)$, respectively. On the other hand, AL routine evaluates QoS violations with complexity of $\mathcal{O}(SK^2)$. Hence, the overall computational complexity can be given as $\mathcal{O}(U(S^2K^3T(K+q))$ for $T$ inner and $U$ outer iterations.
%%%%%%%%%%%%%%%%%%%%%%%%%%%%%%%%%%%%%%%%%%
\vspace{0em}
\section{Numerical Results}
%%%%%%%%%%%%%%%%%%%%%%%%%%%%%%%%%%%%%%%%%%
%%%%%%%%%%%%%%%%%%%%%%%%%%%%%%%%%
In this section, the performance of the D-CAPA metasurface framework with the proposed optimization algorithm is evaluated. We have considered $S \in \{1,\hdots,6\}$ multiple metasurfaces which  serves $K=20$ total users including $L=14$ IUs and $M=6$ EUs.

\begin{figure}[t]
    \centering
    \vspace{-1.8em}
    \includegraphics[trim=10 2 2 0cm,clip,width=0.85\columnwidth]{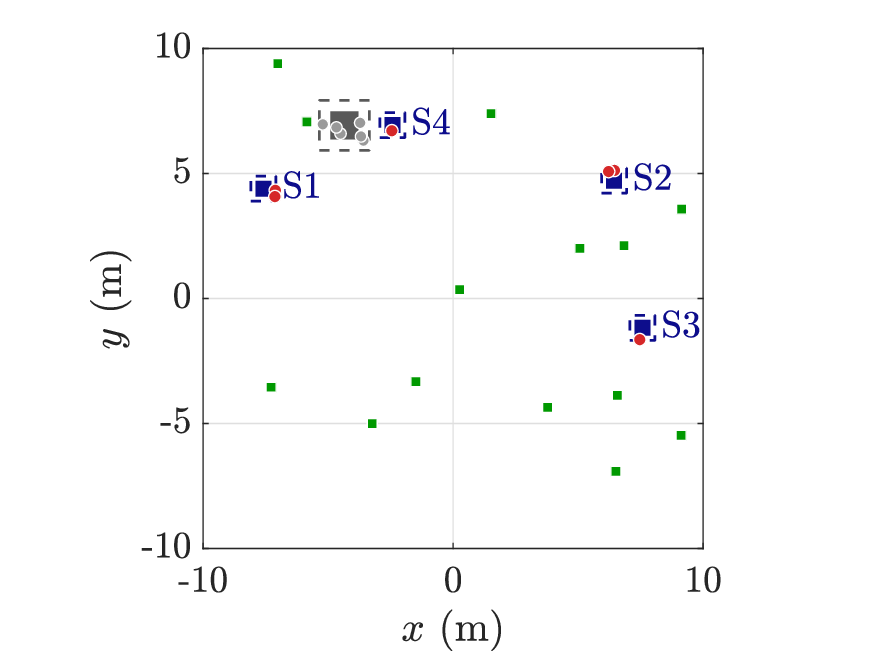}
     \vspace{-0.3em}
    \caption{\small  C‑CAPA and D‑CAPA network layouts (single random realization).\normalsize}
    \vspace{0cm}
    \label{fig:CAPA_layout}

\end{figure}

Figure \ref{fig:CAPA_layout} depicts a random realization of the proposed CAPA metasurfaces framework with D-CAPAs shown as blue squares, while the C-CAPA as grey square. These surfaces are randomly positioned on the $xy$-plane ($x,y \in [-10,10]$ m, $z=0$) with the IUs (green dots) scattered in region $z\in [0.5,20]$ m. The EUs (red dots) are located in small $xy$-areas (dashed blue squares) around the D-CAPAs in region $z\in [0.5,2]$ m. For fair analysis over multiple surface choices, we have considered constant total aperture area $\bar{A}_T=\{0.25,0.5,1\}$ m$^2$ with total system transmission power fixed at $P_t=\{5,10\}$ mA$^2$ \cite{Zhaolin:TCOM_2025}. This corroborates the fact that the same aperture area and transmit power considered for the C-CAPA is further divided into multiple distributed surfaces, $A_T^{(s)}= \bar{A}_T/S$, $P_s= P_t/S$. The system carrier wavelength is fixed at $\lambda=0.1$ m, while the free-space channel impedance $Z_0$ and the EU receiver impedance $Z$ are set to $120 \pi\, \Omega$ and $25\, \Omega$, respectively. The noise power at each user receiver is $\sigma^2=10^{-9}$ A$^2$. The NL-HE parameters are selected as $Q_{max}=24$ mW, while the rectifier parameters as $a=1500$, $b=0.0022$ \cite{Zeeshan:WCL:2025}. The optimization convergence thresholds are fixed as $\delta=10^{-9}$ and $\epsilon=10^{-3}$. For the L-BFGS-B routine, we store only the $q=10$ last curvature pairs for Hessian approximation, while the AL routine sets the initial penalty weight and IU/EU balancing factor as $\rho_0=20$ and $\beta=10$, respectively. 

% %-------------------------
% \begin{figure}[ht]
%     \centering
%     \vspace{-1em}
%     \includegraphics[trim=2cm 1cm 2cm 2cm,clip,width=0.9\columnwidth]{Image/Colocated_CAP_Structure.eps}
%         \vspace{-1em}
%         \caption{\small Single CAP structure.\normalsize}
%         \label{fig:Single_CAP}
% \end{figure}
% %-------------------------

% %-------------------------
% \begin{figure}[ht]
%     \centering
%     \vspace{-1em}
%     \includegraphics[trim=2cm 1cm 2cm 2cm,clip,width=0.9\columnwidth]{Image/Distributed_10_CAP_Structure.eps}
%         \vspace{-1em}
%         \caption{\small Distributed CAP structure.\normalsize}
%         \label{fig:Distributed_CAP}
% \end{figure}
% %-------------------------
%-------------------------
%-------------------------

\begin{figure}[t]
    \centering
    %\vspace{-1em}
    \includegraphics[trim=0cm 0cm 0cm 0cm,clip,width=0.86\columnwidth]{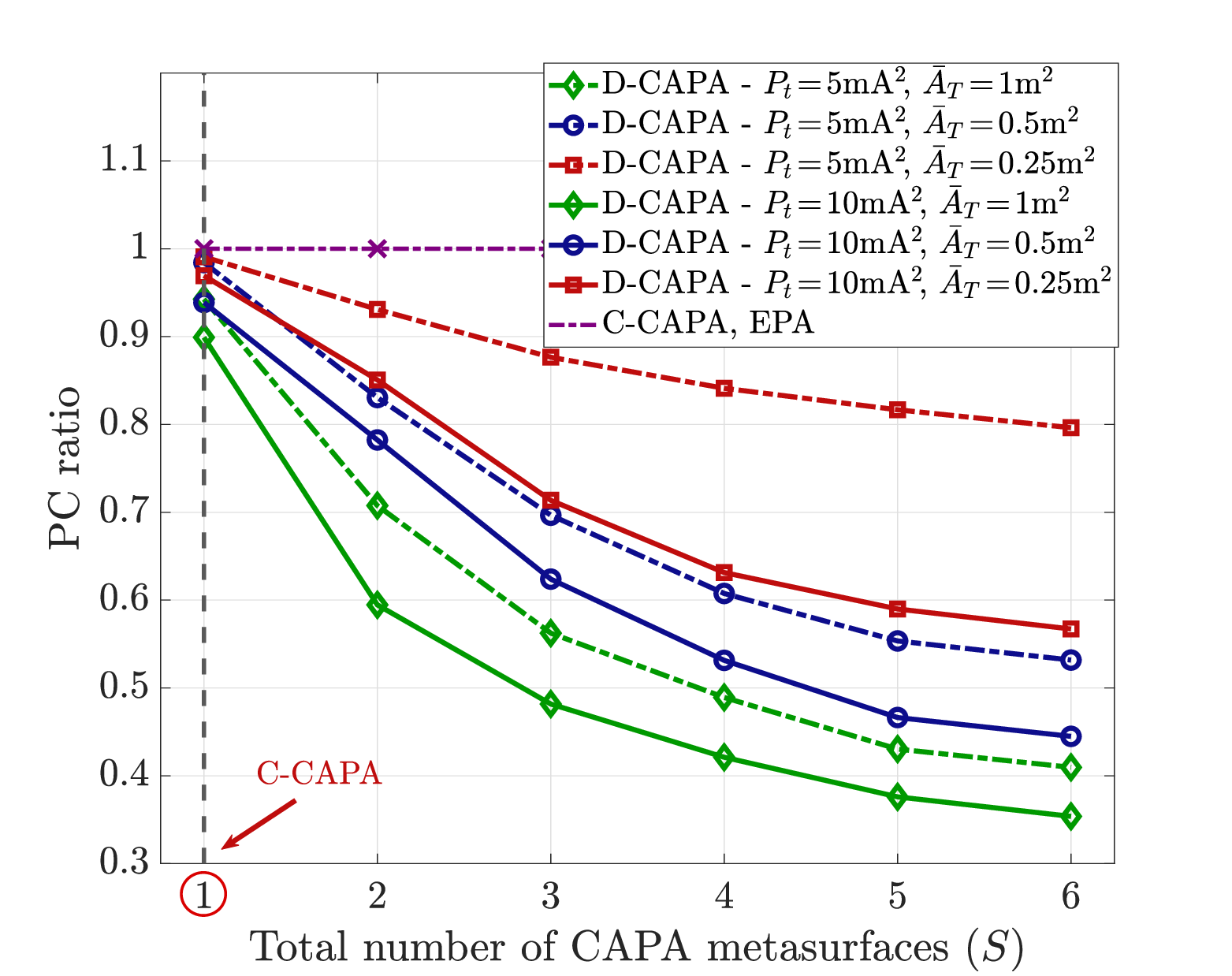}
        \vspace{0em}
        \caption{\small PC ratio versus the number of CAPA surfaces.\normalsize}
        \label{fig:PC_ratio}
        \vspace{0.3em}
\end{figure}

In order to analyze the comparative advantage of the D-CAPA structure with the proposed dual-routine optimization algorithm, we consider the PC ratio against the benchmark \textbf{EPA} case. In this regard, this metric has been analyzed in Fig. \ref{fig:PC_ratio} for multiple number of metasurfaces ($S$) including both C-CAPA ($S=1$) and D-CAPA ($S=\{2,3,4,5,6\}$) cases, while considering different choices of $\bar{A}_t$ and $P_t$. It can be clearly noticed here that the PC ratio reduces progressively with an increasing number of CAPA surfaces, while satisfying the IU SE and EU HE QoS benchmarks. This observation can be attributed to the smaller distances between the CAPA surfaces and the spatially distributed users, even though the aperture area and transmission power for each D-CAPA decreases linearly. For the C-CAPA case, we can observe that our proposed optimization algorithm reduces the PC up to $10\%$ against the non-optimized \textbf{EPA} case (dotted purple line). On the other hand, this reduction is up to $65 \%$ and $61\%$ for the D-CAPA for the $S=6$ scenario with respect to \textbf{EPA} and optimized C-CAPA counterparts, respectively. 

We have also analyzed the effect of different total transmission power $P_t=\{5,10\}$ mA$^2$ (dotted and solid lines), along with the aperture sizes $\bar{A}_T=\{0.25,0.5,1\}$ m$^2$ (red, blue and green lines). Figure \ref{fig:PC_ratio} demonstrate that a larger total aperture size with higher transmission power results in the most effective performance for our proposed optimization algorithm shown by the solid green line. Most importantly, a decreased transmission power has more drastic effect on the PC reduction. For example, we can notice a $40\%$ performance reduction for the case of $P_t=5$ mA$^2$, $\bar{A}_T=0.25$ m$^2$ (dotted red line) in comparison to $P_t=10$ mA$^2$, $\bar{A}_T=0.25$ m$^2$ (solid red line).
%-------------------------
%-------------------------

\begin{figure}[t]
    \centering
    %\vspace{-1em}
    \includegraphics[trim=0cm 0cm 0cm 0cm,clip,width=0.86\columnwidth]{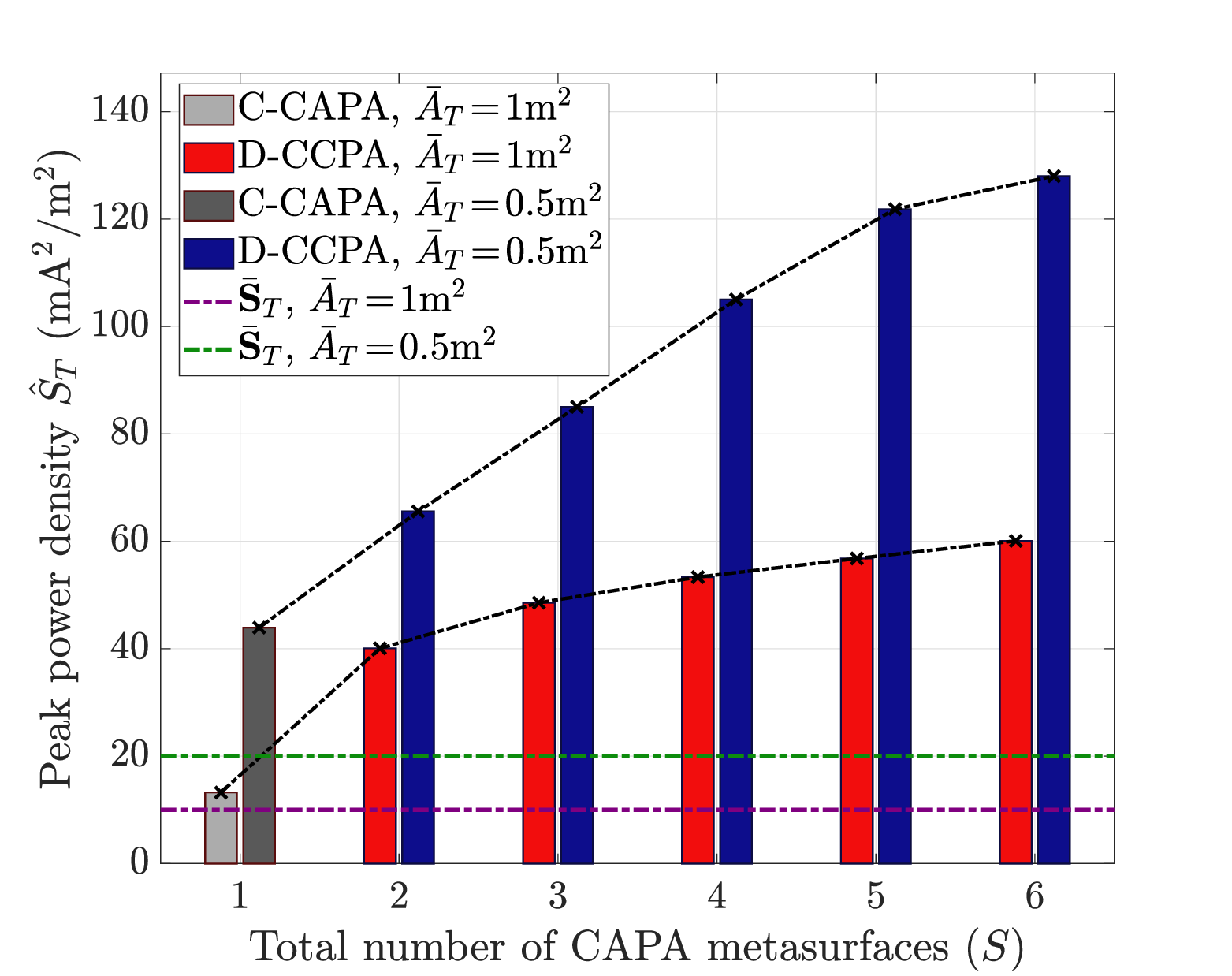}
        \vspace{0em}
        \caption{\small Peak power density versus the number of CAPA surfaces.\normalsize}
        \label{fig:Peak_power_hist}
        \vspace{0.1em}
\end{figure}

Now, we examine the peak transmission power density ($\hat{S}_{T}$) as the number of D-CAPAs increases (colored bars) along with the C-CAPA (grey bars) case  in Fig. \ref{fig:Peak_power_hist}, assuming a constant $P_t=10$ mW$^2$. This analysis also evaluates the impact of the overall CAPA aperture size $\bar{A}_T=\{0.5,1\}$ m$^2$, shown as red and blue bars, respectively.  The proposed dual-routine optimization framework focuses the transmission current distribution by efficiently manipulating the wave amplitudes without compromising on the SE and HE requirements, thus, leading to concentrated energy 'hot-spots' over the metasurfaces. For the C-CAPA case, we observe that $\hat{S}_{T}$ increases by $32\%$ and $120\%$ for $\bar{A}_{T} = 0.5~\text{m}^2$ and $\bar{A}_{T} = 1~\text{m}^2$, respectively, compared with the \textbf{EPA} benchmarks, whose average power densities $\bar{S}_{T}$ are shown as the purple and green dotted horizontal lines. Moreover, the energy focusing becomes substantially stronger in the D-CAPA configuration, achieving up to a six-fold improvement for $S = 6$ compared with the \textbf{EPA} case. This pronounced power concentration over specific regions of the CAPA surface can be attributed to the selective spatial EM visibility across different areas of the metasurface, which guides the beamforming design to concentrate user-specific current distributions onto particular physical zones of the transmitter. Another key observation is that $\hat{S}_{T}$ decreases as the overall aperture size $\bar{A}_{T}$ increases from $0.5~\text{m}^2$ to $1~\text{m}^2$, shown by the red and blue bars, respectively. This implies that for larger apertures, the same transmission power is spread across a wider physical area of the metasurface. 

%-------------------------
%-------------------------

\vspace{-0.5em}
\section{Conclusion}
This paper investigated a distributed CAPA (D‑CAPA) architecture for multiuser SWIPT and developed an optimization framework to minimize system PC under SE and HE QoS constraints derived from the \textbf{EPA} benchmark. By exploiting continuous current‑signal synthesis and dyadic Green’s function–based channels, the proposed framework applies the L‑BFGS‑B method to optimize the PA coefficients and leverages an AL scheme to manage QoS constraint violations. Simulation results demonstrate that the optimized D‑CAPA design achieves up to a $6\%$ reduction in system PC compared with an optimized C‑CAPA configuration, showcasing the strong EM beamforming capabilities of CAPA transmit surfaces. Future work may explore selective user–metasurface association and adaptive activation of D‑CAPA surfaces to further enhance EE.

%==============================================================================
%\vspace{-0.2em}
\bibliographystyle{IEEEtran}
\bibliography{main}
%==============================================================================

\end{document}